# Transfer Learning for Tonal Noise Prediction in VRF Units Using Thermodynamic and Vibration Signals

ZhiWei Su,[1,#] Ding Wang,[1,*] Yuan Guo,[1] Yang Qiao,[1,*] HongJun Cao[1]

[1]Corporate Research Center, Midea Group, Foshan528311, People's Republic of China

## ABSTRACT

*The second-order harmonic (2f) component generated by twin-rotary compressor is a dominant low-frequency noise source of variable refrigerant flow (VRF) outdoor units, yet its amplitude fluctuates strongly with environmental thermal load and valve opening, making it difficult to assess accurately using conventional mechanism-based models. This paper proposes an unsupervised transfer learning method based on Domain-invariant Partial Least Squares (Di-PLS) to accurately predict 2f noise levels under new conditions using different signals. Prediction models utilizing thermodynamic signals and acceleration signals are constructed respectively, and the generalization performance of the proposed Di-PLS is systematically compared with traditional Partial Least Squares (PLS). Results demonstrate that Di-PLS significantly outperforms PLS by extracting cross-condition common features and minimizing the distribution discrepancy between the source and target domains. Specifically, the acceleration-based Di-PLS model achieves the best performance, maintaining prediction errors within 3 dB for all test cases. This superiority over thermodynamic-based models highlights a physical insight: while thermodynamic states drive dynamic changes, structural vibration possesses a stronger and more direct causal link to acoustic radiation.*

## 1. INTRODUCTION

Variable Refrigerant Flow (VRF) multi-split systems are widely used in commercial and residential buildings owing to their high part-load efficiency and precise zonal temperature control. However, the operating noise of VRF outdoor units—particularly the low-frequency components that are difficult to mitigate using conventional passive treatments—has become a key bottleneck for acoustic quality. As the outdoor unit functions as a complex multi-physics coupled system, its noise arises from the interaction of compressor excitation, fan aerodynamic noise, fluid–structure interaction in piping, and refrigerant two-phase flow. For twin-rotary compressors in particular, the characteristic second-order harmonic (2f) component not only dominates the low-frequency energy and strongly affects perceived sound quality [1], but also exhibits large fluctuations with the system thermodynamic state, making single-condition calibration data insufficient for accurate noise assessment in real operation.

In recent years, artificial intelligence (AI) technologies have developed rapidly in heat pump and heating, ventilation, and air-conditioning (HVAC) systems, and have been widely applied to operating condition monitoring, energy-efficiency optimization, thermal comfort control, and fault diagnosis. Hsiao-Ping Ni et al. used deep neural networks for feature enhancement to achieve accurate prediction of temperature and humidity in an air-handling unit, thereby improving HVAC system efficiency [2]; Liu et al. adopted an ensemble reinforcement learning algorithm to coordinately control HVAC and window systems for enhanced overall operating efficiency [3]; Sha et al. proposed a data-driven model predictive control (MPC) strategy to reduce energy consumption while ensuring thermal comfort and indoor air quality [4]; Ren et al. combined thermodynamic prior knowledge with deep learning to efficiently detect sensor faults in large-scale HVAC systems [5]. However, compared with energy-efficiency prediction and optimization or fault diagnosis, research on AI-based acoustic prediction for HVAC systems is still relatively scarce. On one hand, the acquisition and labeling of high-quality acoustic data are costly; on the other hand, the operating conditions of VRF outdoor units are affected by many factors. Under different valve openings and environmental thermal loads, the mapping between the system's thermodynamic state and its acoustic response undergoes significant distribution shifts. Directly applying conventional machine-learning models trained under one operating condition to new conditions often leads to a sharp degradation in generalization performance due to insufficient domain adaptation capability.

[#] Presenting author: suzw28@midea.com
[*] Corresponding author: qiaoyang2@midea.com, ding2.wang@midea.com

To address the above issues, this paper proposes a low-frequency noise prediction method for VRF outdoor units based on unsupervised transfer learning. Using experimental data obtained in a full anechoic chamber under different vapor injection (VI) valve openings and environmental thermal loads, two separate prediction models are constructed with thermodynamic parameters and vibration acceleration signals as inputs. By introducing a transfer-learning mechanism, the goal is to extract transferable feature representations from labeled source-domain conditions and adapt them to completely unlabeled target-domain operating conditions. On this basis, a leave-one-condition-out cross-validation strategy is employed to comparatively analyze the generalization performance and applicability of the two types of physical input signals in fully unlabeled target domains, providing a new approach for low-cost online monitoring of low-frequency noise from VRF outdoor units.

## 2. DATA ACQUISITION AND PREPROCESSING

To build a dataset combining thermodynamic, vibration, and acoustic information, synchronous measurements were carried out on a VRF outdoor unit in a full anechoic chamber. The test unit has a rated cooling capacity of 18 kW and is driven by a twin-rotary vapor-injection compressor, whose injection mass flow can be finely adjusted via the injection valve openings, enabling flexible switching of the compressor thermodynamic state. A Siemens SCADAS Mobile system was used for vibration and noise signal acquisition. According to the company's internal standard, eight microphones were positioned at a distance of 1 m around the unit at equal height, forming a circular array (Figure 1 (a), N1–N8). For structural vibration, 39 acceleration channels were instrumented using a combination of single-axis and triaxial accelerometers. Sensor locations covered key vibration transmission paths and radiating surfaces, including the compressor shell, suction and discharge lines, and the outdoor-unit panels (Figure 1 (b) shows the accelerometer layout on the discharge side). In parallel, 66 high-precision thermocouples were installed to monitor the entire refrigerant loop, covering critical thermodynamic nodes such as the suction and discharge ports and mid-sections of the evaporator and condenser flow passages. All thermocouple junctions were tightly wrapped with insulation material to minimize the influence of ambient thermal radiation on temperature measurements (Figure 1 (c), suction side layout).

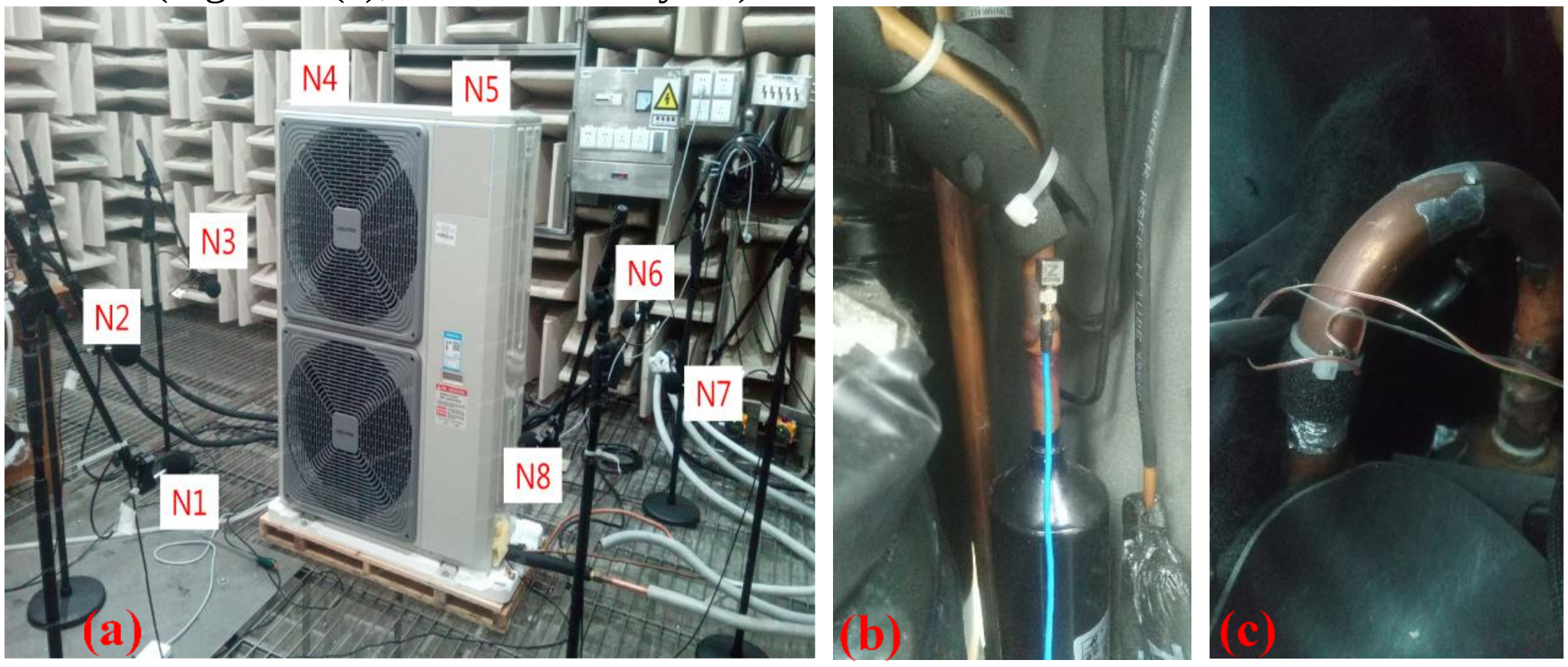

Figure 1: Schematic diagram of the experimental setup and typical sensor arrangement.

The test operating condition was set to the rated heating mode. The compressor was kept running at its rated frequency, with the main electronic expansion valve opening fixed at its nominal value.

The vapor injection valve opening was adjusted to change the thermodynamic cycle state of the system. A set of operating conditions covering various injection states was designed, and each condition was maintained for 1–2 hours after the system reached thermal equilibrium. During the experiment, the wet-bulb temperature in the anechoic chamber naturally fluctuated between −1 °C and 5 °C, which changed the heat load of the outdoor unit and dynamically affected the thermodynamic state of the system. Accordingly, this environmental factor was recorded simultaneously as an independent variable. Data acquisition was triggered at 1-minute intervals, capturing a continuous 10-second stream per event. Vibration and acoustic signals were sampled at 20 kHz, and signal averaging was applied to enhance the signal-to-noise ratio. For quasi-static thermodynamic parameters such as temperature, the arithmetic mean over each 10-second window served as the representative feature value for the corresponding sample.

The present study focuses on predicting the second-order harmonic (2f) component associated with the compressor operating frequency. The data preprocessing procedure was as follows. First, a Fast Fourier transform (FFT) was applied to the raw vibration and noise signals, and the root-mean-square (RMS) value within a ±3 Hz band around the 2f frequency was extracted and converted into decibels (dB). Second, the label of each sample was defined as the arithmetic mean of the sound pressure level at 2f over the eight microphones, while the input features consisted of the 2f amplitudes from the 39 acceleration channels and the 66 thermodynamic temperature measurements at the corresponding time. In total, 19 representative operating conditions with different vapor injection valve openings (denoted Case 1 to Case 19) were tested. After discarding non-steady-state data during condition switching and samples affected by sensor anomalies, 1492 valid samples remained. The number of samples per condition ranged from 40 to 120, and the 2f noise labels spanned from 40 dB to 55 dB (range of 15 dB). As shown in Figure 2, due to the combined effects of ejector-valve adjustment and ambient temperature fluctuation, the data distributions of different condition groups exhibit pronounced and irregular statistical shifts. It is worth noting that the dataset contains two special cases: Case 4 is an “automatic regulation” condition in which the valve opening dynamically varies between 90 and 105 according to the internal control logic, and Case 5 is a “valve-closed” condition in which the vapor-injection circuit is deactivated, while all other cases have fixed openings. The inclusion of these special conditions further increases the complexity and distribution discrepancy of the dataset.

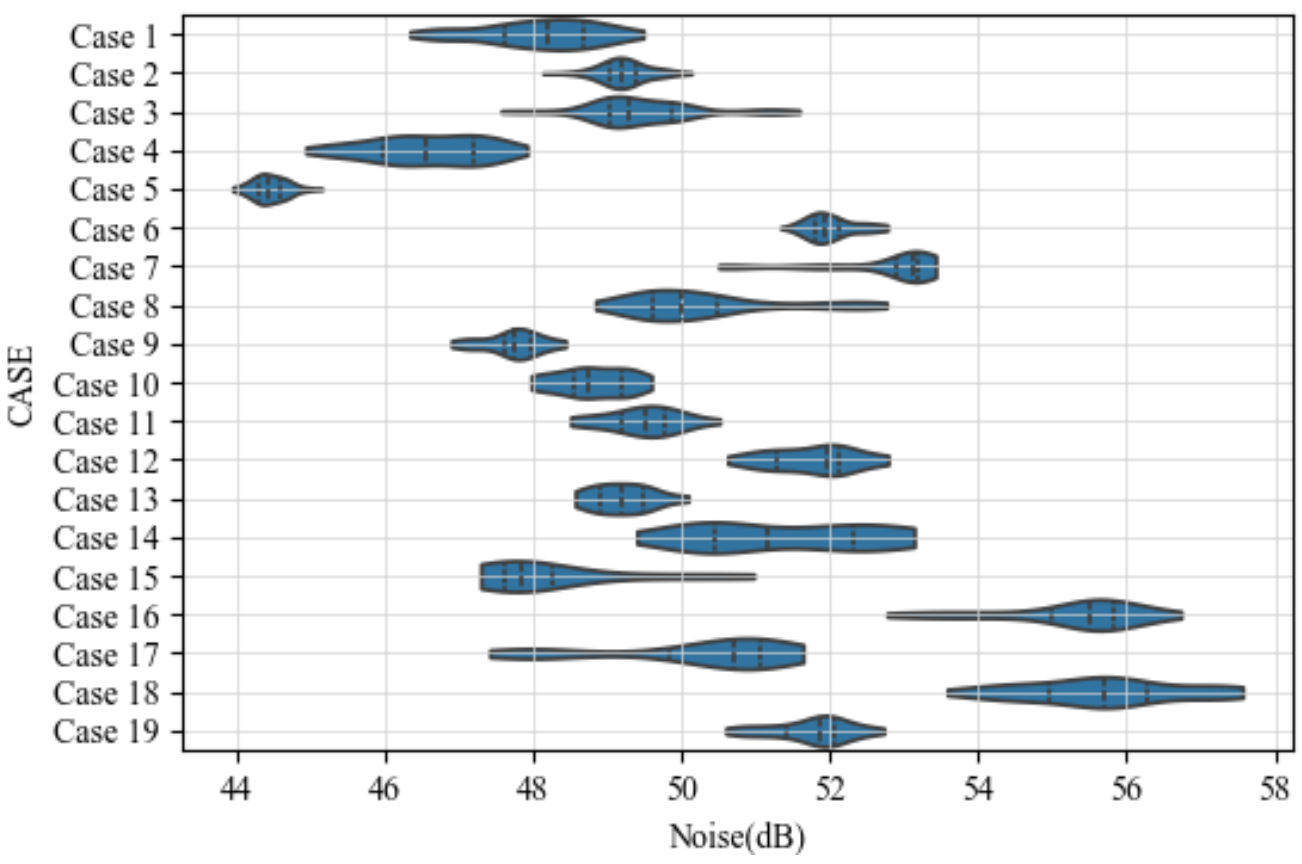

Figure 2: Statistical distribution of 2f noise level samples under different operating conditions.

Furthermore, to visualize the data distribution characteristics, we employed the Uniform Manifold Approximation and Projection (UMAP) dimensionality reduction technique [6], projecting both vibration and thermodynamic features onto a 2D plane, as shown in Figure 3. The left panel displays the UMAP projection for vibration signals, and the right panel for thermodynamic parameters. In both panels, data points are color-coded by valve opening condition. These UMAP visualizations confirm significant covariate shift across operating conditions. Data points from different conditions (distinct colors) do not exhibit continuous transitions but instead aggregate into clearly isolated clusters. This pronounced separation and lack of overlap between clusters signify a form of covariate shift, highlighting the necessity for domain adaptation strategies. Both feature types manifest this clustered structure, indicating that each operating condition forms a distinct manifold in the feature space.

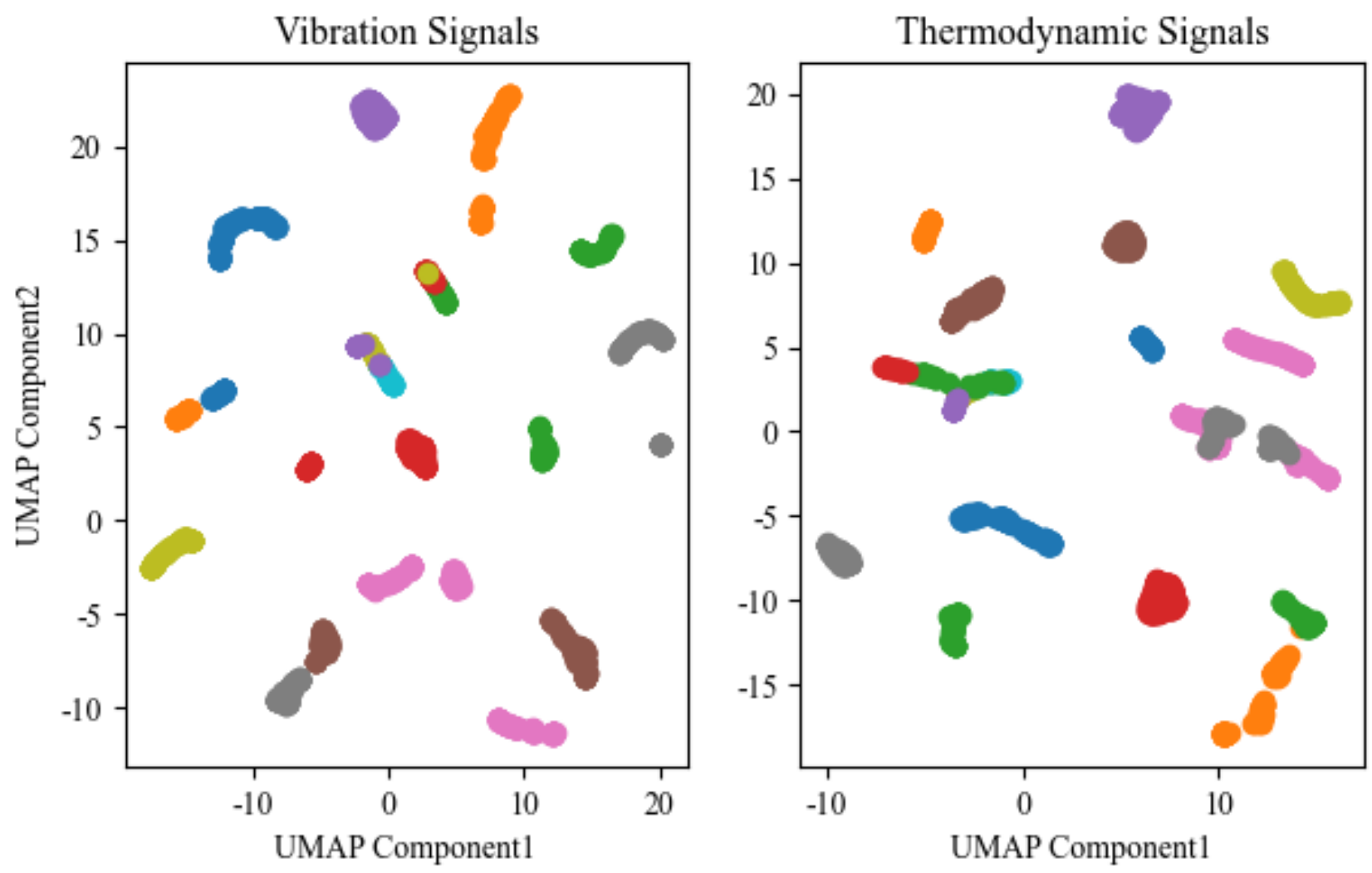

Figure 3: UMAP projection of vibration signals and thermodynamic signals.

## 3. METHODOLOGY

Partial Least Squares (PLS) regression has been widely used for performance prediction of industrial systems due to its strong modeling capability under small sample sizes and highly correlated inputs [7]. PLS searches for linear combinations of the original predictor matrix $\boldsymbol{X}$ and response vector $\boldsymbol{y}$ that maximize the covariance between them. For a single latent variable, the optimization objective can be formulated as:

$$\boldsymbol{\omega}_{\mathrm{PLS}} = \underset{\boldsymbol{w}}{\operatorname{argmin}} \left\| \boldsymbol{X} - \boldsymbol{y}\boldsymbol{\omega}^{\mathrm{T}} \right\|_F^2 \tag{1}$$

Its solution can be:

$$\boldsymbol{\omega}_{\mathrm{PLS}}^{\mathrm{T}} = \frac{\boldsymbol{y}^{\mathrm{T}}\boldsymbol{X}}{\boldsymbol{y}^{\mathrm{T}}\boldsymbol{y}} \tag{2}$$

However, the conventional PLS algorithm relies on the assumption that the training and test sets are independently and identically distributed. When facing covariate shift where the input distributions of the source and target domains are inconsistent (e.g., significant thermodynamic state shifts caused by changes in vapor injection valve opening), the latent space constructed from the source domain cannot be effectively adapted to the target-domain data, leading to a sharp degradation of the model's generalization performance on unlabeled, unknown operating conditions (target domain).

To mitigate this limitation, the present study adopts the domain-invariant partial least squares (Di-PLS) method [8]. The key idea of Di-PLS is to augment the original PLS objective with a domain-alignment regularization term, which explicitly minimizes the statistical distribution discrepancy between the source and target domains in the latent space, thereby enforcing the model to learn domain-invariant and transferable features. The objective function of Di-PLS is defined as:

$$\boldsymbol{\omega}_{di-PLS} = \underset{\boldsymbol{w}}{\operatorname{argmin}} \left\| \boldsymbol{X} - \boldsymbol{y}\boldsymbol{\omega}^{\mathrm{T}} \right\|_F^2 + \lambda \left| \frac{1}{n_{\mathrm{S}} - 1} \boldsymbol{\omega}^{\mathrm{T}} \boldsymbol{X}_{\mathrm{S}}^{\mathrm{T}} \boldsymbol{X}_{\mathrm{S}} \boldsymbol{\omega} - \frac{1}{n_{\mathrm{T}} - 1} \boldsymbol{\omega}^{\mathrm{T}} \boldsymbol{X}_{\mathrm{T}}^{\mathrm{T}} \boldsymbol{X}_{\mathrm{T}} \boldsymbol{\omega} \right| \tag{3}$$

Here, $\lambda \geq 0$ serves as the regularization parameter that balances the trade-off between predictive performance and domain invariance, $n_{\mathrm{S}}$ and $n_{\mathrm{T}}$ denote the numbers of samples in the source and target domains, respectively. A key advantage of this method lies in the existence of a closed-form solution. By taking the gradient of Eq. (3) with respect to $\boldsymbol{\omega}$ and setting it to zero, the corresponding weight vector can be obtained as:

$$\boldsymbol{\omega}_{di-PLS}^{\mathrm{T}} = \frac{\boldsymbol{y}^{\mathrm{T}}\boldsymbol{X}}{\boldsymbol{y}^{\mathrm{T}}\boldsymbol{y}}\left[\mathbf{I} + \frac{\lambda}{2\boldsymbol{y}^{\mathrm{T}}\boldsymbol{y}}\left(\frac{1}{n_{\mathrm{S}}-1}\boldsymbol{X}_{\mathrm{S}}^{\mathrm{T}}\boldsymbol{X}_{\mathrm{S}} - \frac{1}{n_{\mathrm{T}}-1}\boldsymbol{X}_{\mathrm{T}}^{\mathrm{T}}\boldsymbol{X}_{\mathrm{T}}\right)\right]^{-1} \tag{4}$$

# 4. MODEL EVALUATION AND ANALYSIS

To rigorously assess the model's generalization performance under unseen operating conditions, a leave-one-condition-out cross-validation scheme is employed. Specifically, each operating condition is treated in turn as the target domain for testing, and the prediction results from all folds are subsequently aggregated for a unified performance assessment. In addition to conventional metrics such as mean squared error (MSE) and coefficient of determination ($R^2$), this paper specifically introduces indicators that are of greater practical engineering value for heat pump systems—namely, the proportions of samples whose absolute prediction errors are less than 2 dB and 3 dB (denoted as $\mathrm{Acc}_{<2\mathrm{dB}}$ and $\mathrm{Acc}_{<3\mathrm{dB}}$ in this paper). These indicators are designed to quantitatively assess the reliability of predictions under actual engineering tolerance requirements, thereby ensuring that the prediction results provide practical guidance.

## 4.1. Acceleration-based model

Using acceleration signals as inputs, the PLS and Di-PLS algorithms are respectively employed, and the scatter plots comparing the predicted values with the test values are shown in Figure 4. The black dashed line in the figure represents the ideal 1:1 prediction line, while the light gray shaded area corresponds to the ±3 dB error band. To ensure fairness of comparison and optimal model performance, the numbers of latent variables for both models are determined via cross-validation and fixed at 14, so that the algorithm performance is evaluated under the same model size.

A comparison of the two panels reveals that the prediction points produced by the PLS model are more widely scattered around the reference line than those of the Di-PLS model. This discrepancy is particularly evident under operating conditions with substantial distribution mismatch, such as the valve-closing condition in Case 5, where the prediction errors become markedly larger. Constrained by the distribution shift across operating conditions, the overall predictive accuracy of PLS is relatively poor, achieving an $R^2$ of only 0.56, an MSE of 3.98 dB, and $\mathrm{Acc}_{<2\mathrm{dB}}$ and $\mathrm{Acc}_{<3\mathrm{dB}}$ of 79.7% and 86.8%, respectively. By contrast, the Di-PLS model in the right panel delivers substantially better predictive performance: sample points from nearly all conditions are tightly concentrated around the diagonal, and even for those challenging scenarios where PLS essentially breaks down, Di-PLS still maintains a strong fit. Quantitative results show that the $R^2$ of Di-PLS increases to 0.86, the MSE drops to 1.46 dB, and $\mathrm{Acc}_{<2\mathrm{dB}}$ and $\mathrm{Acc}_{<3\mathrm{dB}}$ rise to 91.4% and 99.6%, respectively. These findings confirm that incorporating a transfer-learning strategy effectively mitigates the data distribution shift of the VRF outdoor unit induced by mode switching and environmental variability, and markedly enhances both the accuracy and robustness of cross-condition prediction.

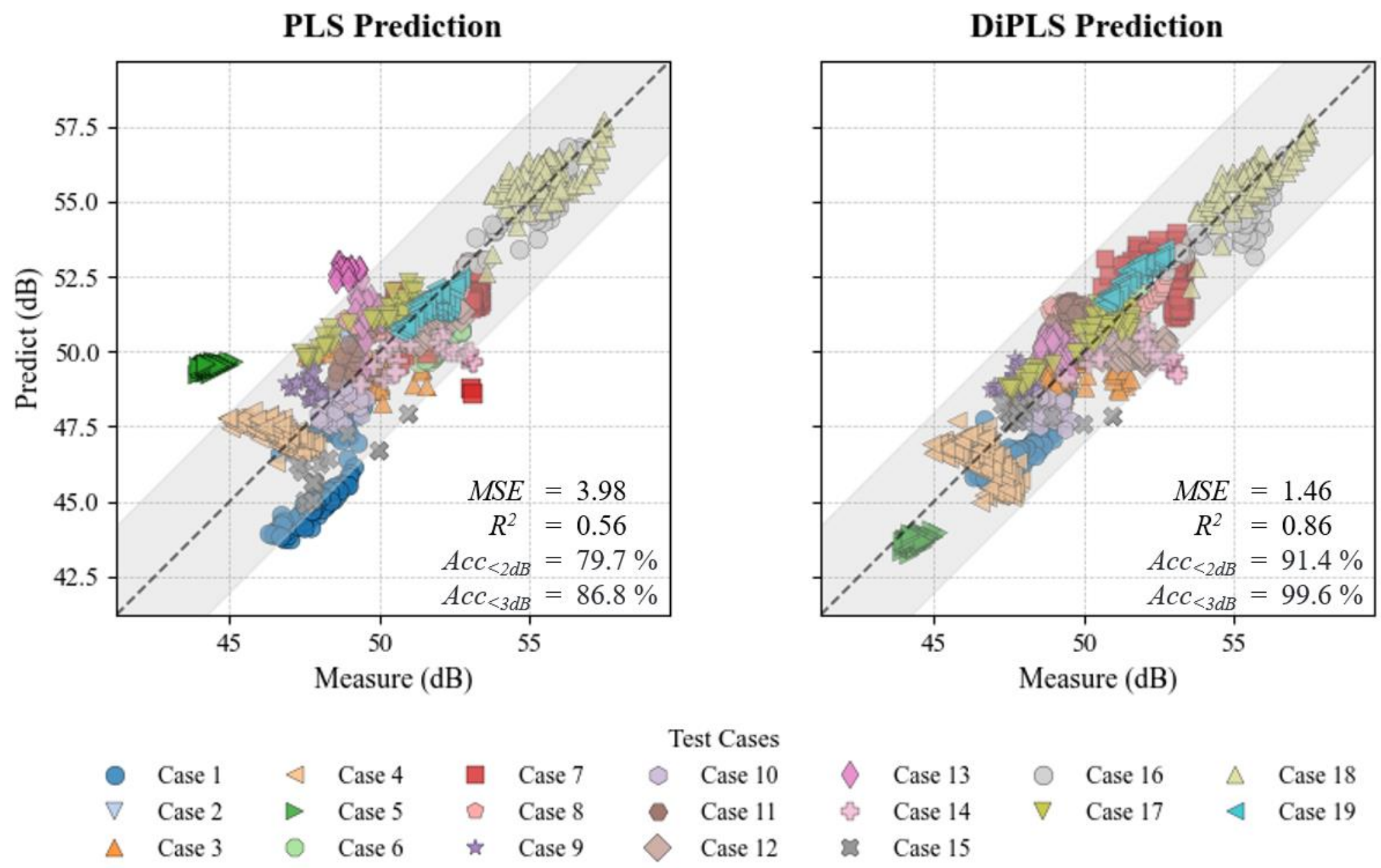

Figure 4: Scatter plots of predicted vs. actual noise levels for different models: (left) PLS model; (right) Di-PLS model.

To elucidate the mechanism underlying the superior performance of Di-PLS, the "valve closed" condition (Case 5), which differs markedly from the other operating conditions, is selected for detailed analysis. Under this condition, the vapor injection circuit is deactivated, and the thermodynamic cycle on the compressor side becomes completely different from that of the conventional valve-open condition. Regarding predictive accuracy, Di-PLS exhibits extremely high adaptability under this specific condition, with a prediction MSE of only 1.55 dB, which is much lower than the 25.74 dB obtained by conventional PLS. Figure 5 shows the projected distributions of the source and target domain data in the first two latent variable spaces of the PLS and Di-PLS models, respectively. The yellow and green regions represent the distribution ranges of samples from the source and target domains, respectively. It can be observed that, after the feature mapping of Di-PLS, the target-domain sample points are completely enclosed by those from the source domain, whereas the PLS method does not exhibit such a complete inclusion relationship in the latent space, and there remain disjoint regions between the two domains. To further quantify the extent of distribution alignment, the Wasserstein distance was employed as a metric. This distance effectively measures the geometric discrepancy between the data distributions of the source and target domains; the smaller its value, the closer the target-domain features are to those of the source domain [9]. Quantitative analysis reveals that the Wasserstein distances corresponding to the Di-PLS and PLS methods are 12.1 and 45.4, respectively. Therefore, it can be concluded that by incorporating domain alignment through Di-PLS, the distance between the source and target domains is significantly reduced, enabling the feature distribution of the target domain to be successfully aligned and mapped into the source domain space, which in turn leads to a substantial improvement in prediction accuracy.

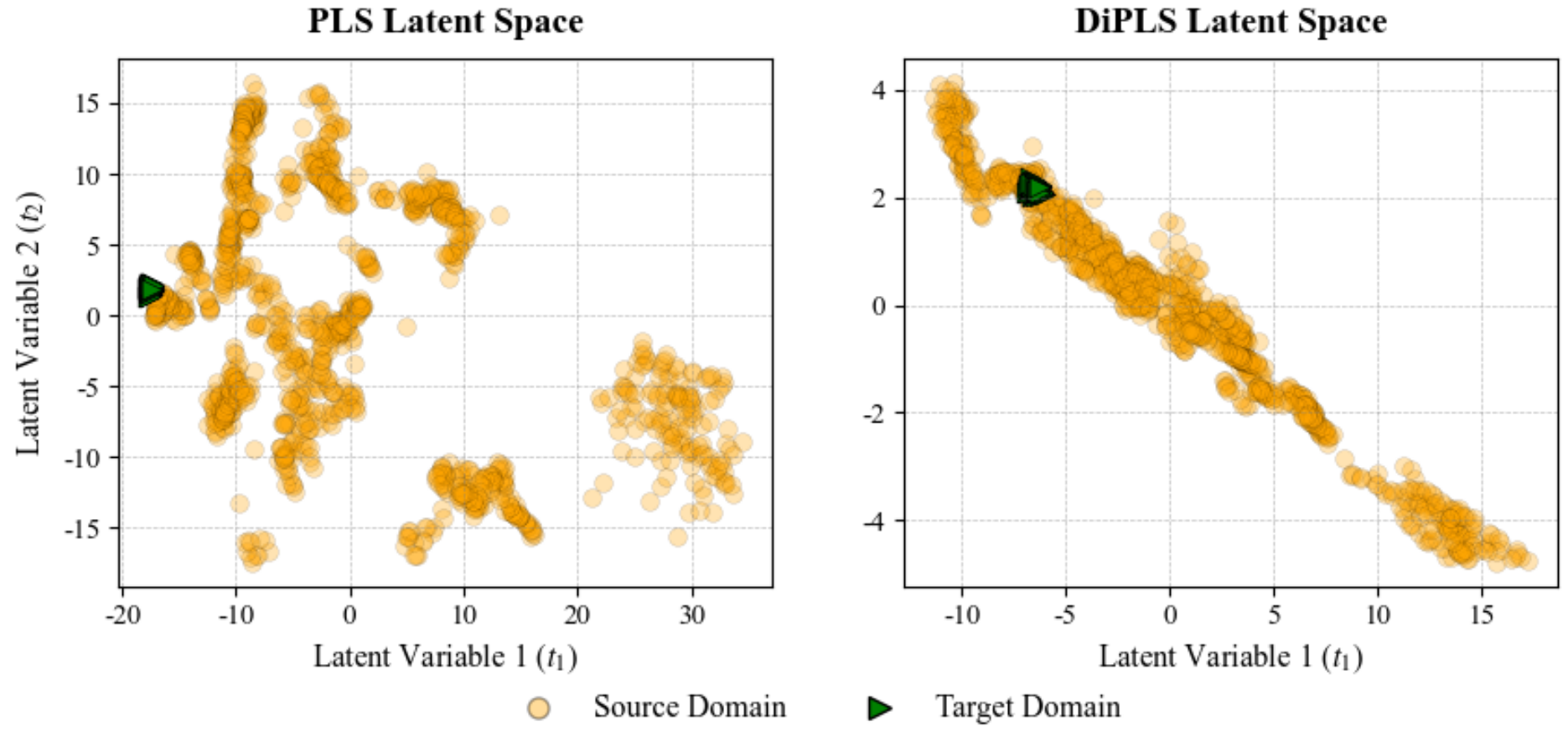

Figure 5: Latent variable space distributions of source and target domains for PLS and Di-PLS models with the injection valve closed.

## 4.2. Thermodynamics-based model

Figure 6 presents the noise prediction results using thermodynamic parameters as inputs, with the operating condition labels consistent with Figure 4. To examine the impact of special operating conditions on model performance, two comparative settings are considered: the right panel shows the prediction results after removing Case 4 ("automatic regulation") and Case 5 ("valve closed") from the dataset, whereas the left panel includes all operating conditions, i.e., including Case 4 and Case 5. A comparison between Figure 4 and the left panel of Figure 6 first indicates that the overall predictive accuracy of the thermodynamic-parameter-based model is inferior to that of the acceleration-based model. Furthermore, a side-by-side comparison of the left and right panels in Figure 6 shows that the inclusion of Case 4 and Case 5 leads to a pronounced deterioration in model performance. When these two operating conditions are excluded (right panel), the model still retains a moderate predictive capability (MSE = 3.03 dB, $R^2$ = 0.51). Once they are included (left panel), however, the MSE surges to 12.12 dB, $R^2$ drops to 0.28, and $Acc_{<2dB}$ and $Acc_{<3dB}$ plummet from 70.0% and 95.4% to 34.3% and 48.2%, respectively.

This outcome highlights the limitations of thermodynamic parameters as inputs in the presence of severe concept drift. In Case 4 ("automatic regulation"), the real-time dynamic adjustment of the valve introduces pronounced non-stationarity, leading to persistent fluctuations in system thermodynamic variables and a shift in the noise generation mechanism from quasi-steady flow behavior to highly dynamic transient behavior. In Case 5 ("valve closed"), the vapor injection circuit is deactivated, and the compressor cycle changes from a dual-suction to a single-suction configuration. This not only reconfigures the internal flow paths and thermodynamic states of the compressor, but also fundamentally alters the noise generation mechanism by eliminating injection-induced flow pulsations. Such mechanistic changes induce a substantial mismatch between the source- and target-domain distributions, and a limited set of thermodynamic parameters cannot adequately represent this near-fundamental change in the noise mechanism. By contrast, the superior accuracy achieved by the acceleration-based model under identical conditions further corroborates a more intrinsic causal coupling between vibration signals and noise radiation. Unlike thermodynamic parameters, which are indirect and highly sensitive to operating conditions, vibration signals are more capable of capturing the essential characteristics of acoustic radiation, thereby enabling more robust and reliable noise prediction.

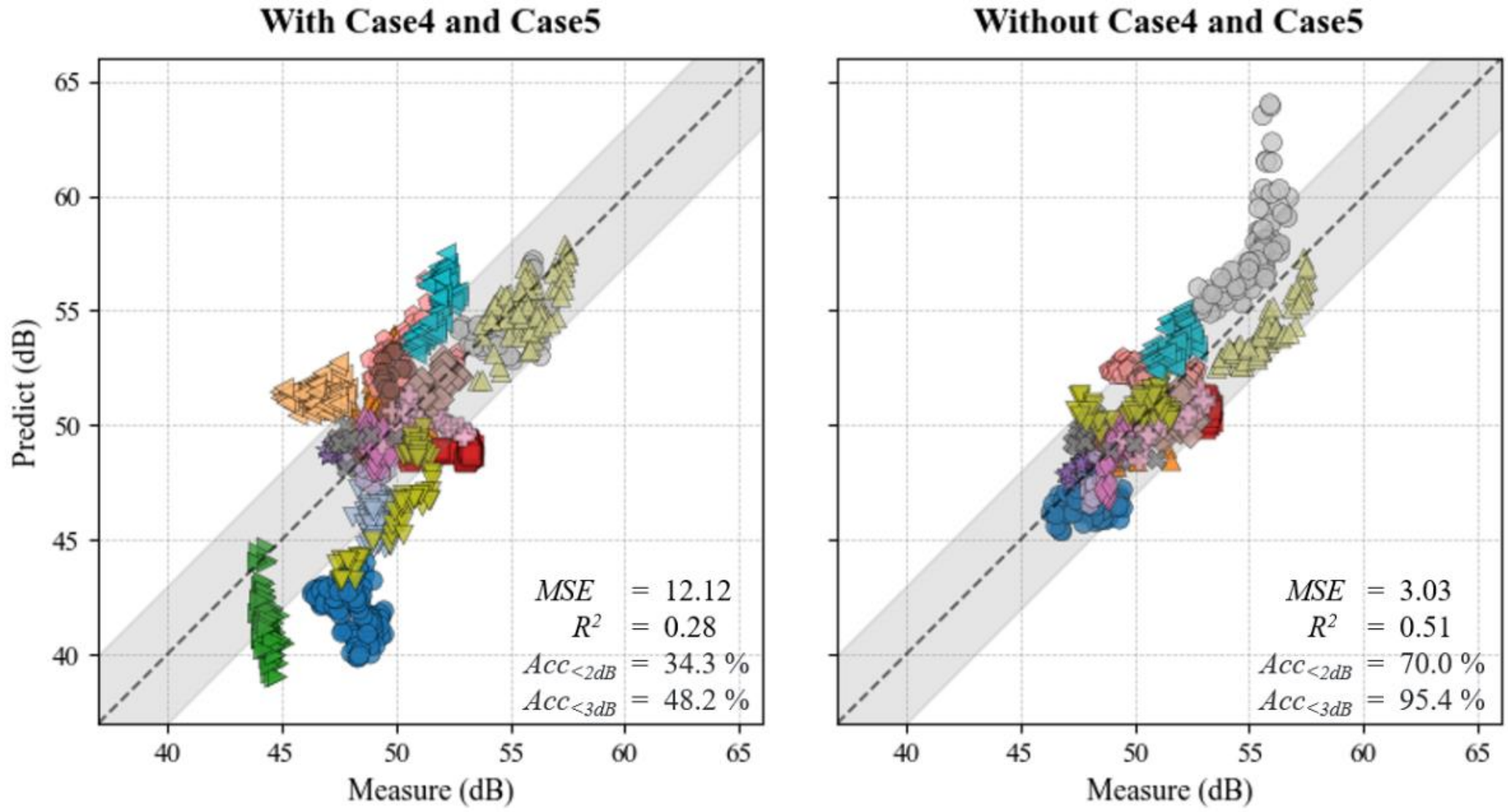

Figure 6: Comparison of noise prediction results based on thermodynamic parameters under different operating condition combinations.

## 5. CONCLUSION

In response to the need for low-frequency noise prediction of VRF air-conditioning systems under multi-condition operation, this study constructs a prediction model using the Di-PLS transfer learning algorithm. The results show that this algorithm can effectively extract domain-invariant features across different operating conditions, and its generalization performance is significantly better than that of conventional PLS under the same model complexity. Further feature analysis reveals that, compared with indirectly related thermodynamic parameters, vibration signals with strong physical causality exhibit superior robustness under changing operating conditions, ensuring that prediction errors for nearly all samples remain within the ±3 dB tolerance. This study verifies the effectiveness of the proposed transfer learning framework for noise prediction of VRF outdoor units, and provides a reliable modeling strategy for acoustic evaluation of heat pump air-conditioning systems under complex operating conditions.